\documentclass[final,5p,twocolumn]{elsarticle}
\usepackage{l3regex}
\usepackage{lipsum}
\usepackage[numbers]{natbib}
\usepackage{graphicx}
\usepackage{hyperref}
\usepackage{dcolumn}
\usepackage{amssymb}
\usepackage{amsmath}
\journal{Physics Letters B}
\begin{document}

\begin{frontmatter}

  \title{PandaX-II Constraints on Spin-Dependent WIMP-Nucleon Effective Interactions}

  \author[1]{Jingkai Xia}
  \address[1]{School of Physics and Astronomy, Shanghai Jiao Tong University, Shanghai Key Laboratory for Particle Physics and Cosmology, Shanghai 200240, China}
  \author[2]{Abdusalam Abdukerim}
  \address[2]{School of Physics and Technology, Xinjiang University, \"{U}r\"{u}mqi 830046, China}
  \author[1]{Wei Chen}
  \author[1]{Xun Chen}
  \author[3]{Yunhua Chen}
  \address[3]{Yalong River Hydropower Development Company, Ltd., 288 Shuanglin Road, Chengdu 610051, China}
  \author[1]{Xiangyi Cui}
  \author[4]{Deqing Fang}
  \address[4]{Shanghai Institute of Applied Physics,  Chinese Academy of Sciences, Shanghai 201800, China}
  \author[1]{Changbo Fu}
  \author[1]{Karl Giboni}
  \author[1]{Franco Giuliani}
  \author[1]{Linhui Gu}
  \author[3]{Xuyuan Guo}
  \author[5]{Zhifan Guo}
  \address[5]{School of Mechanical Engineering, Shanghai Jiao Tong  University, Shanghai 200240, China}
  \author[1]{Ke Han}
  \author[1]{Changda He}
  \author[3]{Shengming He}
  \author[1]{Di Huang}
  \author[6]{Xingtao Huang}
  \address[6]{School of Physics and Key Laboratory of Particle Physics and Particle Irradiation
    (MOE), Shandong University, Jinan 250100, China}
  \author[1]{Zhou Huang}
  \author[7]{Peng Ji}
  \address[7]{School of Physics, Nankai University, Tianjin 300071, China}
  \author[1,8]{Xiangdong Ji\fnref{fn1}}
  \ead{xdji@sjtu.edu.cn}
  \address[8]{Tsung-Dao Lee Institute, Shanghai 200240, China}
  \author[5]{Yonglin Ju}
  \author[1]{Shaoli Li}
  \author[1]{Heng Lin}
  \author[5]{Huaxuan Liu}
  \author[1,8]{Jianglai Liu}
  \author[4]{Yugang Ma}
  \author[9]{Yajun Mao}
  \address[9]{School of Physics, Peking University, Beijing 100871,China}
  \author[1]{Kaixiang Ni}
  \author[3]{Jinhua Ning}
  \author[1]{Xiangxiang Ren}
  \author[1]{Fang Shi}
  \author[10]{Andi Tan}
  \address[10]{Department of Physics, University of Maryland, College Park, Maryland 20742, USA}
  \author[6]{Anqing Wang}
  \author[5]{Cheng Wang}
  \author[4]{Hongwei Wang}
  \author[6]{Meng Wang}
  \author[4]{Qiuhong Wang}
  \author[9]{Siguang Wang}
  \author[5]{Xiuli Wang}
  \author[1]{Xuming Wang}
  \author[5]{Zhou Wang}
  \author[7]{Mengmeng Wu}
  \author[3]{Shiyong Wu}
  \author[10,11]{Mengjiao Xiao}
  \address[11]{Center of High Energy Physics, Peking University, Beijing 100871, China}
  \author[8]{Pengwei Xie}
  \author[6]{Binbin Yan}
  \author[1]{Jijun Yang}
  \author[1]{Yong Yang}
  \author[7]{Chunxu Yu}
  \author[6]{Jumin Yuan}
  \author[3]{Jianfeng Yue}
  \author[10]{Dan Zhang}
  \author[1]{Hongguang Zhang}
  \author[1]{Tao Zhang}
  \author[1]{Li Zhao}
  \author[12]{Qibin Zheng}
  \address[12]{School of Medical Instrument and Food Engineering, University of Shanghai for Science and Technology, Shanghai 200093, China}
  \author[3]{Jifang Zhou}
  \author[1]{Ning Zhou\corref{cor1}}
  \ead{nzhou@sjtu.edu.cn}
  \author[9]{Xiaopeng Zhou}
  
  \author[13,14]{Wick C. Haxton\corref{cor2}}
  \address[13]{Department of Physics, University of California, Berkeley, CA 94720, USA}
  \address[14]{Lawrence Berkeley National Laboratory, Berkeley, CA 94720, USA}
  \ead{haxton@berkeley.edu}
  
  \fntext[fn1]{Spokesperson of PandaX Collaboration}
  \cortext[cor1]{Corresponding author}
  \cortext[cor2]{Corresponding author}

  \begin{abstract}
    We present PandaX-II constraints on candidate WIMP-nucleon effective interactions involving the nucleon or WIMP spin, including, in addition to standard axial spin-dependent (SD) scattering, various couplings among vector and axial currents, magnetic and electric dipole moments, and tensor interactions. The data set corresponding to a total exposure of 54-ton-days is reanalyzed to determine constraints as a function of the WIMP mass and isospin coupling. We obtain WIMP-nucleon cross section bounds of $\rm 1.6 \times 10^{-41} cm^2$ and $\rm 9.0 \times 10^{-42} cm^2$  ($90\%$ c.l.) for neutron-only SD and tensor coupling, respectively, for a mass $M_\mathrm{WIMP} \sim {\rm 40~GeV}/c^2$.  The SD limits are the best currently available for $M_\mathrm{WIMP} > {\rm 40~GeV}/c^2$.   We show that PandaX-II has reached a sensitivity sufficient to probe a variety of other candidate spin-dependent interactions at the weak scale.
    
  \end{abstract}
  \begin{keyword}
    PandaX-II experiment \sep WIMP dark matter \sep Spin-dependent effective interactions
  \end{keyword}
  
\end{frontmatter}

Astrophysical and cosmological observations indicate that a large amount of non-luminous dark matter (DM) exists in our universe, constituting $\sim27\%$ of the closure density. However, the exact nature of DM remains a mystery. One intriguing DM candidate, a weakly-interacting massive particle (WIMP), arises naturally in many extensions of the standard model~\cite{feng2010,bertone2005}. Many WIMP searches have been performed, including direct detection of their scattering off target nuclei, indirect detection of their decay or annihilation, and their production in collider experiments.  In the analysis of direct detection experiments, {\color{black} frequently it is assumed that the scattering arises from the light-quark-level ($u$, $s$, $d$) effective interaction
\begin{equation}
\mathcal{L} \sim {G_F \over \sqrt{2}}\sum_q  \left[ \bar{\chi} \gamma_\mu \chi ~c_q^{VV} \bar{q} \gamma^\mu q + \bar{\chi} \gamma_\mu \gamma_5 \chi ~c_q^{AA} \bar{q} \gamma^\mu \gamma_5 q \right] 
\label{eq:neutralino}
\end{equation}
which can be reduced to a nucleon-level operator useful in analyzing the nuclear response to WIMP scattering \cite{klos2013}. 
Limits on the vector spin-independent (SI) and axial spin-dependent (SD) WIMP-proton and WIMP-neutron cross sections $\sigma_{p,n}^\mathrm{SI,SD}$
can then be derived.  The recent stringent direct detection null results
obtained successively by LUX~\cite{lux2017,luxsd2017}, PandaX~\cite{pandax2017,pandaxsd} and XENON~\cite{xenon2017,xenonsd,xenonrecent} have significantly tighten the bounds
on $\sigma_{p,n}^\mathrm{SI,SD}$~\cite{jianglai2017}.

The interaction of Eq. (\ref{eq:neutralino}) was motivated by supersymmetric DM candidates, like the neutralino, that can naturally
account for the DM relic density.  The motivation to focus exclusively on such candidates has weakened due in particular 
to collider constraints~\cite{peskin2015}.  An alternative approach,
effective field theory (EFT)~\cite{liantao2010,katz2013,haxton2014}, has gained favor because it allows one to do an analysis~\cite{cdmseft,xenoneft,zuowei2017,gaurav2018} free of
theory assumptions.  One selects an EFT scale -- e.g., the light-quark or a nucleon
scale -- and constructs a complete basis of effective operators to a given order, taking into account all general symmetries 
limiting that basis.   The underlying UV theory of DM will reduce at that scale to some definite combination of the basis
operators, regardless of its nature.  Experimentalists can explore the sensitivities of their detectors to the basis operators, to make
sure they are probing all possibilities.  The EFT approach has shown 1) relative experimental sensitivities depend on the operator choice,
and 2) direct detection is potentially more powerful than might appear 
from SI/SD analyses, as six (not two) independent constraints on DM can be obtained, in principle~\cite{haxton2014}.}

The PandaX-II detector, located in the China Jinping Underground Laboratory (CJPL), 
is a dual-phase xenon time-projection chamber with 580~kg of liquid xenon in the sensitive target volume. When the incoming WIMP scatters off a xenon nucleus, both the prompt scintillation photons (S1) in the liquid and the delayed proportional scintillation photons (S2) in the gas are collected by 55 top and 55 bottom Hamamatsu R11410-20 3-inch photomultiplier tubes. 
The experiment has an accumulated exposure of 54-ton-days~\cite{pandax2017}. 
The previously reported analysis for the standard (isoscalar) SI interaction yielded a 90\% exclusion limit on $\sigma^\mathrm{SI}$ of $8.6\times 10^{-47} \rm cm^2$ 
for a WIMP mass of $40~{\rm GeV}/c^2$.  (These bounds were recently superseded by XENON1T results \cite{xenonrecent}.)
{\color{black} In this paper we present PandaX-II constraints a variety of candidate interactions that depend on nucleon
or WIMP spin.

Spin-dependent interactions other than SD can arise from WIMP magnetic and electric dipole moments,  vector-axial interference terms, tensor interactions, etc.  While the associated 
nucleon-level effective operators are conventionally expressed in covariant
form, they can be rewritten in terms of the Galilean-invariant EFT basis used here, convenient for nonrelativistic shell model (SM) treatments of the nuclear physics.  This basis consists of fourteen operators generated at next-to-next-to-leading-order from the
nucleon-WIMP perpendicular relative velocity operator $\vec{v}^{\perp}$, the momentum transfer $\vec{q}$, and the WIMP and nucleon spins, $\vec{S}_{\chi}$ and  $\vec{S}_{N}$ \cite{haxton2014}.}

We specialize to the scattering of a spin-1/2 WIMPs off a natural xenon target, exploring
four dimension-four and  three higher dimension effective interactions, selected from Table 1 of Ref.~\cite{haxton2014}.  The operator dimension is defined as 4 + number of powers of $m_M$ in the denominator, where $m_M$ is a scale that governs the strength of the WIMP and nucleon moments being coupled.  The dimension-four
operators are the V-A interactions
\begin{align}
  \mathcal{L}_{\rm{int}}^5 \equiv \bar{\chi}\gamma^{\mu}\chi\bar{N}\gamma_{\mu}N
  \rightarrow   \mathcal{O}_1 \nonumber
\end{align}
\begin{align}
  \mathcal{L}_{\rm{int}}^7 \equiv \bar{\chi}\gamma^{\mu}\chi\bar{N}\gamma_{\mu}\gamma^5 N
  \rightarrow  -2\mathcal{O}_7+2\frac{m_N}{m_{\chi}}\mathcal{O}_9 \nonumber
\end{align}
\begin{align}
  \mathcal{L}_{\rm{int}}^{13} \equiv \bar{\chi}\gamma^{\mu}\gamma^5\chi\bar{N}\gamma_{\mu} N
  \rightarrow  2\mathcal{O}_8+2\mathcal{O}_9 \nonumber
\end{align}
\begin{align}
  \mathcal{L}_{\rm{int}}^{15} \equiv \bar{\chi}\gamma^{\mu}\gamma^5\chi\bar{N}\gamma_{\mu}\gamma^5N
   \rightarrow  -4\mathcal{O}_4 \ .
\end{align}
with $m_N$ the nucleon mass and  $\mathcal{O}_1=1_{\chi}1_N $, $\mathcal{O}_4=\vec{S}_\chi \cdot \vec{S}_N$, $\mathcal{O}_7 = \vec{S}_N \cdot \vec{v}^\perp$, $\mathcal{O}_8 = \vec{S}_\chi \cdot \vec{v}^\perp$, and $\mathcal{O}_9=i (\vec{S}_\chi \times
\vec{S}_N) \cdot {\vec{q} \over m_N}$ the nonrelativistic operators of \cite{katz2013,haxton2014}.  While  $\mathcal{L}_{\rm{int}}^5$
generates the standard SI interaction, the other interactions involve an axial coupling and thus depend on spin.  

One can equally well start with a basis of light-quark effective operators, reducing these via chiral EFT to their nonrelativistic nucleon equivalents
\cite{hoferichter2015,grinstein2017}.  The spin-dependent nucleon-level operators arising from the axial part of  Eq. (1) (the standard SD interaction) and from the light-quark tensor 
interaction will be considered here.

The dimension-five operators coupling the WIMP magnetic or
electric dipole moments with the nucleon's
vector current, and the dimension-six operator coupling WIMP and nucleon magnetic moments, are examples of other
potential sources of spin-dependent scattering,
\begin{flalign}
 ~& \mathcal{L}_{\rm{int}}^9  \equiv \bar{\chi}i\sigma^{\mu\nu}\frac{q_{\nu}}{m_M}\chi\bar{N}\gamma_{\mu}N& \nonumber\\
  & \rightarrow  -\frac{\vec{q}^{\, 2}}{2m_{\chi} m_M}\mathcal{O}_1+\frac{2m_N}{m_M}\mathcal{O}_5-\frac{2m_{N}}{m_M}(\frac{\vec{q}^{\, 2}}{m_{N}^2}\mathcal{O}_4-\mathcal{O}_6) & \nonumber
\end{flalign}
\begin{flalign}
~ & \mathcal{L}_{\rm{int}}^{17} \equiv  i\bar{\chi}i\sigma^{\mu\nu}\frac{q_{\nu}}{m_M}\gamma^5\chi\bar{N}\gamma_{\mu}N
  \rightarrow \frac{2m_N}{m_M}\mathcal{O}_{11}& \nonumber
  \label{eqn:L17}
\end{flalign}
\begin{flalign}
~  \mathcal{L}_{\rm{int}}^{10} &\equiv  \bar{\chi}i\sigma^{\mu\nu}\frac{q_{\nu}}{m_M}\chi\bar{N}i\sigma_{\mu\alpha}\frac{q^{\alpha}}{m_M} N
 \rightarrow  4(\frac{\vec{q}^{\, 2}}{m_M^2}\mathcal{O}_4-\frac{m_N^{ 2}}{m_M^2}\mathcal{O}_6)&
\end{flalign}
Here $\mathcal{O}_5=i\vec{S}_{\chi}\cdot({\vec{q} \over m_N}\times\vec{v}^{\perp})$, $\mathcal{O}_6=(\vec{S}_{\chi}\cdot {\vec{q} \over m_N})(\vec{S}_{N}\cdot {\vec{q} \over m_N})$, and $\mathcal{O}_{11}=i\vec{S}_{\chi} \cdot {\vec{q} \over m_{N}}$.  The 
dependence on $\vec{q}$ implies nuclear form factors that peak at larger momentum transfers, influencing
experimental strategies for optimally constraining such interactions.  We set $m_M \equiv m_N$, normalizing both WIMP and nucleon
moments to the nucleon scale.

Each operator $\mathcal{L}^i_\mathrm{int}$ can have independent couplings to protons and neutrons, or equivalently to isospin \cite{haxton2014},
\begin{equation}
{d_i^0 \over m_V^2} + {d_i^1 \over m_V^2} \tau_3 =  {d_i^p \over m_V^2}{1 + \tau_3 \over 2} + {d_i^n \over m_V^2}{1- \tau_3 \over 2}
\label{eq:ds}
\end{equation}
where $\tau_3$ is the Pauli isospin operator.  The  couplings $d_i$ are dimensionless, defined relative to the weak scale
\[ m_V \equiv \langle v \rangle = (2 G_F)^{-{1 \over 2}} = 246.2 ~\mathrm{GeV} \]
with $\langle v \rangle$ the Higgs vacuum expectation value. We will consider isoscalar ($d_i^1=0$) and isovector ($d_i^0=0$) interactions,
for which $d_i^p=d_i^n$ and  $-d_i^p=d_i^n$, respectively; as well as
couplings only to protons ($d_i^1= d_i^0$) or neutrons ($d_i^1=- d_i^0$).  For our higher-dimension operators, the $d_i$'s
also encode information about the absolute size of the WIMP electric and dipole moments, given that we have normalized to nucleonic values
($m_M\equiv m_N)$.

As the spin response of Xe is largely governed by the unpaired neutrons in $^{129,131}$Xe, one expects
the $1-\tau_3$ projection of nucleon-spin operators to dominate.

\begin{figure}[htbp]
  \centering
  \includegraphics[width=0.45\textwidth]{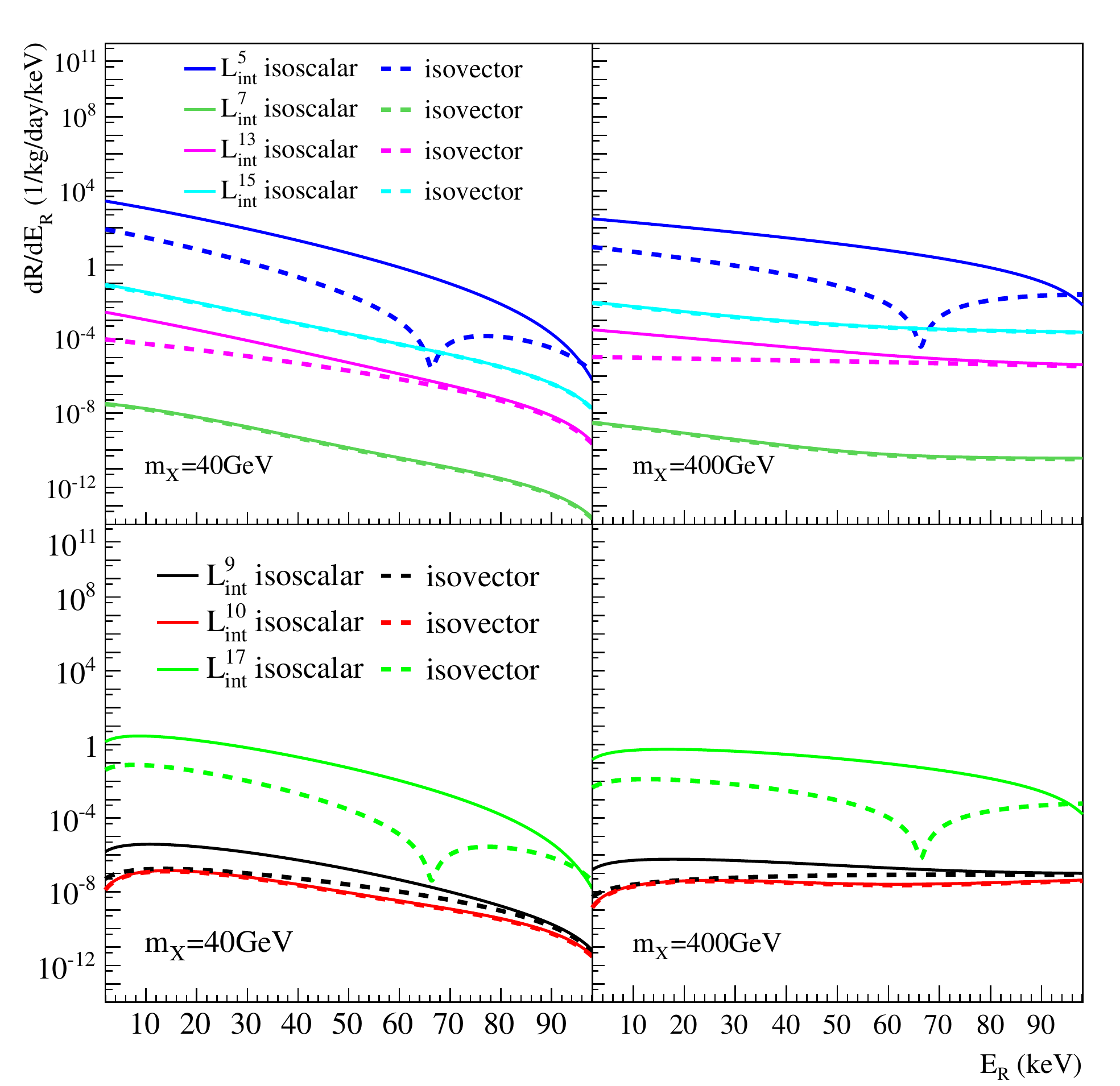}
  \caption{Recoil energy spectra for the scattering of spin-${1 \over 2}$ WIMPs on xenon nuclei, for WIMP masses of ${\rm 40~GeV}/c^2$ and ${\rm 400~GeV}/c^2$, and
  unit isoscalar ($d_i^0=1$, $d_i^1=0$) and isovector ($d_i^0=0$, $d_i^1=1$) coupling.  Top panel: dimension-four V-A interactions.  Bottom panel: higher dimension magnetic and
  electric dipole interactions.}
  \label{fig:spectra_EFT}
\end{figure}

The differential cross section for elastic scattering is
\begin{eqnarray}
  \frac{d\sigma(v,E_R)}{dE_R} &=& 2 m_T   \frac{d\sigma(v, \vec{q}^{\,2})}{d \vec{q}^{\,2}} ~~~~\nonumber \\
  &=&   \frac{2m_T}{4\pi v^2}\left [\frac{1}{2J_{\chi}+1}\frac{1}{2J+1}\sum_{\rm spins}\left|\mathcal{M}\right|^2\right ]~~~~
\end{eqnarray}
where the square of the Galilean invariant amplitude $\mathcal{M}$ is a product of WIMP and nuclear matrix elements and is a function of initial WIMP velocity $v$ (dimensionless, in units of $c$) and the three-momentum transfer $\vec{q}$ \cite{haxton2014}.  Here $J_\chi$ is the WIMP spin and $J$ the nuclear
ground state angular momentum.  In the long wavelength limit the nuclear response functions corresponding to matrix elements
of simple operators familiar from weak interactions, such as $1(i)$, $\vec{\sigma}(i)$, $\vec{\ell}(i)$, and
$\vec{\sigma}(i) \cdot \vec{\ell}(i)$.
The differential event rate with respect to nuclear recoil energy is
\begin{equation}
  \frac{dR}{dE_R} = \frac{\rho_{\chi}}{m_{\chi}}\int{\frac{d\sigma(v,E_R)}{dE_R} v f(\vec{v})d^3v},
  \label{eqn:velocity}
\end{equation}
where the $f(\vec{v})$ is the normalized velocity distribution of the WIMP particles.
We calculate WIMP signal rates by evaluating Eq. (\ref{eqn:velocity}) for a local WIMP mass density of $\rho_{\chi}=0.3~{\rm GeV}/c^2/{\rm cm^3}$,
assuming a Maxwellian WIMP velocity distribution peaked at $v_0=220~\rm km/s$ and truncated at the galactic escape velocity $v_{esc}=544~\rm km/s$.

The nuclear response functions for DM elastic scattering must be calculated before experimental limits can be
converted to bounds on the operator coefficients of Eq. (\ref{eq:ds}).   For each contributing Xe isotope and needed operator, we performed full-basis shell-model calculations
using the GCN5082~\cite{GCN5082} interaction
(so named because the SM valence space resides between the shell closures at nucleon numbers 50 and 82).
The calculations were done without truncation, using the SM code BIGSTICK to treat bases that ranged to 9 billion
Slater determinants~\cite{bigstick}. Full GCN5082 Xe isotopes response functions are used in the Mathematica script
of \cite{haxton2018}, an update of \cite{haxton2014}.  This script and the associated library of one-body nuclear
density matrices are available on request from the authors of \cite{haxton2018}.   The scripts of \cite{haxton2018} and \cite{haxton2014} were carefully
cross-checked against one another, to verify their consistency.   

Figure~\ref{fig:spectra_EFT} shows computed recoil energy spectra.  The upper panels include the coherent isoscalar  (N+Z) and isovector (N-Z)  SI responses (operator $\mathcal{L}^5_\mathrm{int}$),
which we show for normalization.  Though we used SM results, simple phenomenological forms~\cite{lewin1996} also work well, below the
diffraction minimum, as the form factor is governed by its known $q=0$ value and the nuclear radius.

The remaining curves corresponding to interactions with nucleon spin  ($\mathcal{L}^7_\mathrm{int}$), WIMP spin
($\mathcal{L}^{13}_\mathrm{int}$), or both ($\mathcal{L}^{15}_\mathrm{int}$), at two WIMP masses.   All curves correspond to weak-scale interactions:
with $d_i^p = d_i^n \equiv 1$ ($d_i^p =-d_i^n \equiv 1$) for isoscalar (isovector) coupling. 
The lower panels show the
corresponding results for the magnetic and electric dipole moment interactions,
$\mathcal{L}^9_\mathrm{int}$, $\mathcal{L}^{10}_\mathrm{int}$, and $\mathcal{L}^{17}_\mathrm{int}$.

The WIMP couplings $d_i$ are constrained by PandaX-II rate limits.  
We use data from two low-background physics runs with a total exposure of 54-ton-days, Run~9 with 79.6 live days in 2016 and Run~10 with 77.1 live days in 2017.
Calibration data from an AmBe source outside the cryogenic vessel and tritium decays from $\rm CH_3 T$ injected into the Xe provided tests of the detector response to the 
nuclear (NR) and electron (ER) recoil events, respectively~\cite{pandax2017}. A NEST-based Monte Carlo (MC) simulation (Ref.~\cite{nestER} for ER, Ref.~\cite{nestNR} for NR) of the data in the (S1, S2) distribution is optimized by adjusting the initial excitation-to-ionization ratio and the recombination fluctuation. A reliable tuned NEST model for PandaX-II is obtained for S1 up to 50 
photoelectrons (PEs). The S1 and S2 signal distributions for a given EFT interaction are simulated using the tuned NEST model and PandaX-II detector response parameters~\cite{pandax2017}.

Our event selection criteria follow Ref.~\cite{pandax2017}: S1 from 3 to 45~PEs, S2 from 100 (raw, not corrected for electron lifetime) to 10000~PEs, events lying within the 99.99\% NR acceptance, and the total fiducial target of $329\pm16$~kg. The backgrounds in Run~9 and Run~10, estimated in Ref.~\cite{pandax2017}, are dominated by $\rm ^{127}Xe$, tritium, other flat ER background ($\rm ^{85}$Kr, radon and detector gamma background), accidental, and neutron contributions.

This search window can include nuclear recoil event energies up to $\rm 100~keV_{nr}$ (NR energy) due to the smearing of S1 and S2, although the efficiency drops below 50\% above $\rm 35~keV_{nr}$ on average. For dimension-four operators, the majority of signal events have nuclear recoil energy below 35~$\rm keV_{nr}$. The overall signal selection efficiency is between 40\% and 50\%, very similar to that for the SI analysis in Ref.~\cite{pandax2017}. For dimension-five or dimension-six operators, the analysis is more complicated due to the sharper form factor
momentum dependence. For instance, the signal efficiency
for the isovector $\mathcal{L}^{9}_\mathrm{int}$ operator decreases from 65\% to 6.5\% as the WIMP mass increases from 40 to 400~GeV. {\color{black} Thus efficiencies can be improved by adjusting 
search windows according to operator type and WIMP mass, as LUX has described \cite{Larsen2018}. Such strategies will be explored in future PandaX analyses: detector calibration studies for S1 above 50~PE will be needed to implement such window adjustments.}

\begin{figure}[htbp]
  \center
  \includegraphics[width=0.45\textwidth]{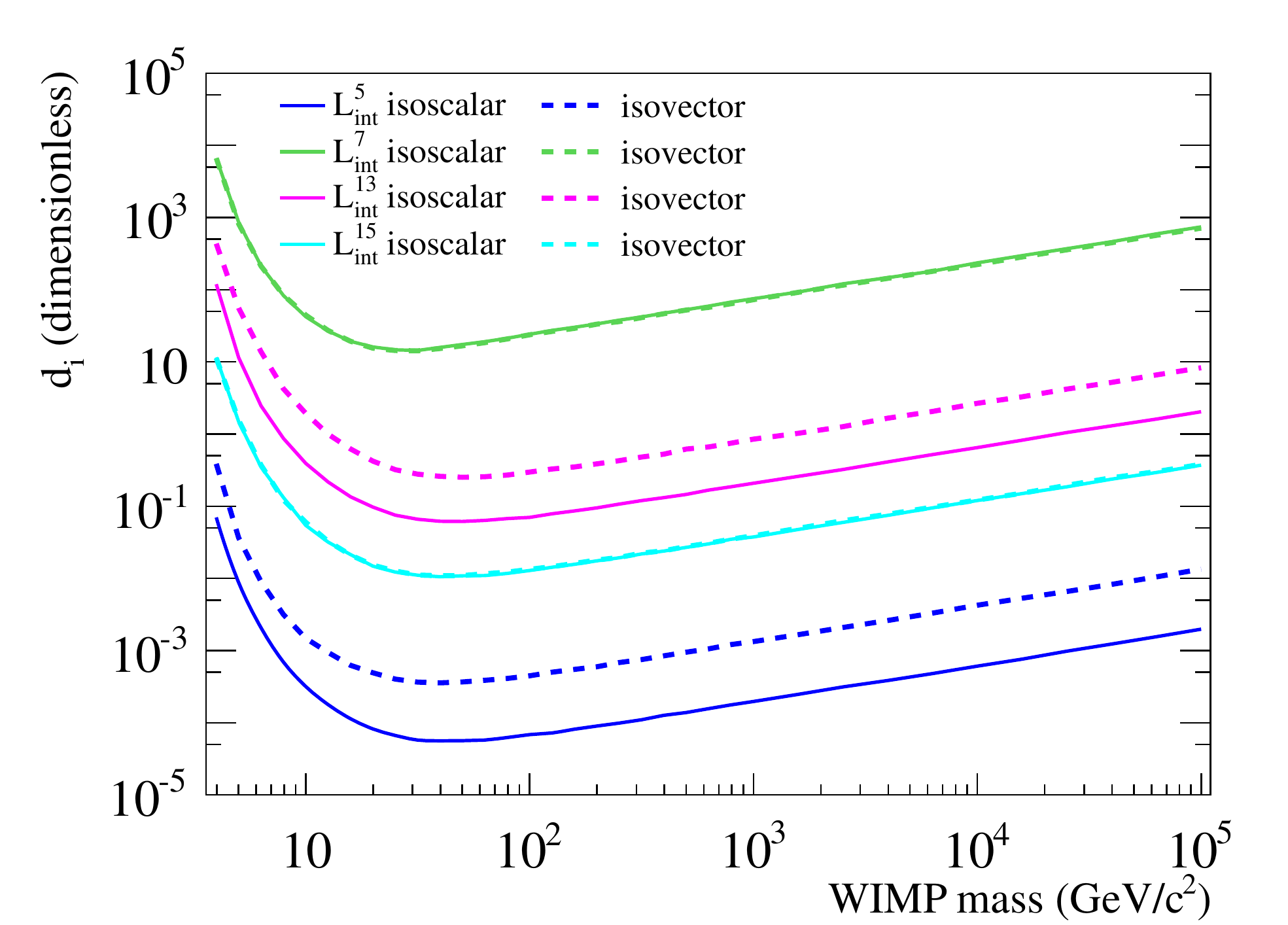}
  \includegraphics[width=0.45\textwidth]{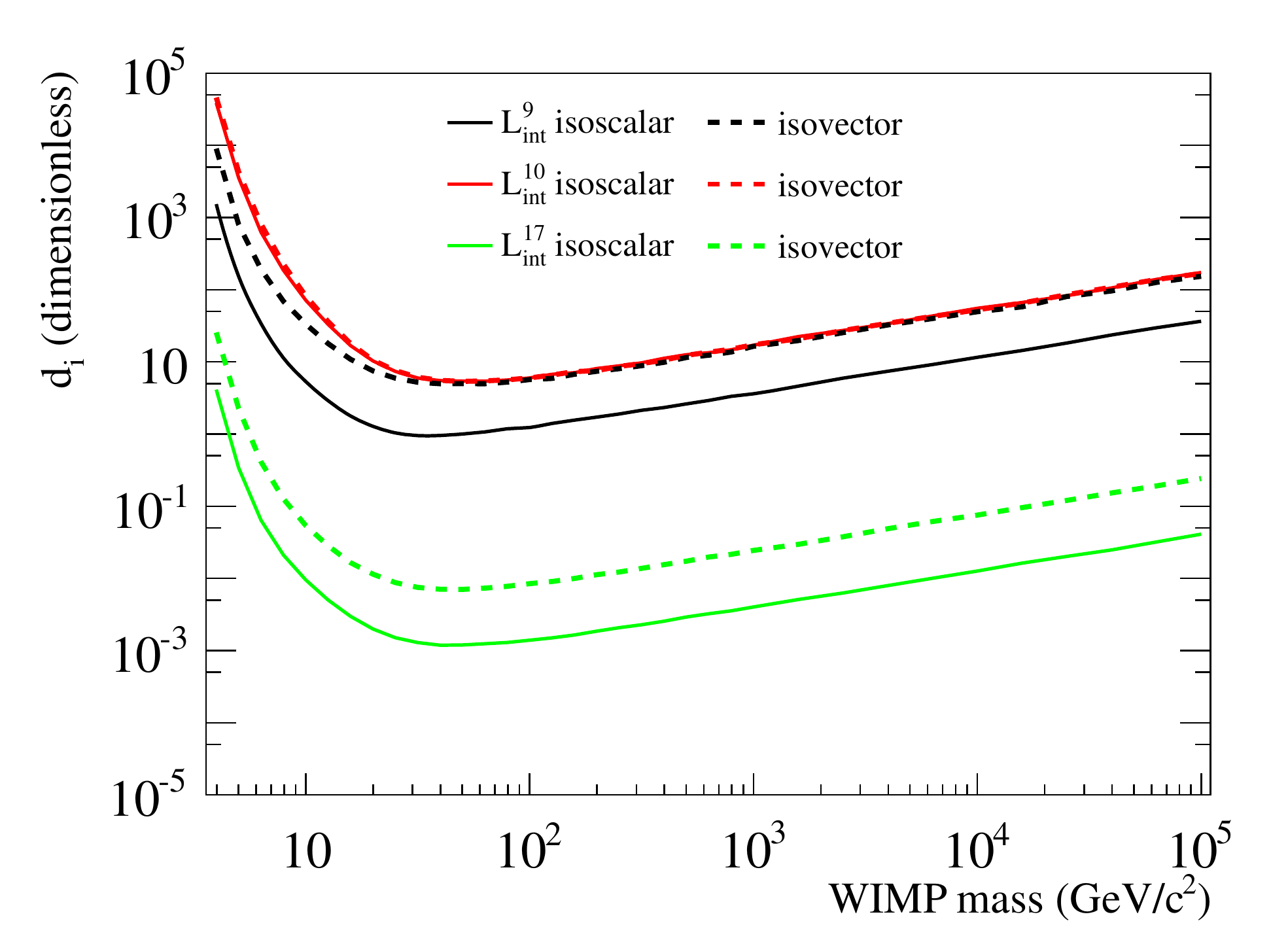}
  \caption{Exclusion limits on the coefficients of dimension-four (top panel) and higher dimension (bottom panel) operators.}
  \label{fig:EFT}
\end{figure}

We constrain the WIMP-nucleon EFT couplings $d_i$ as a function of the WIMP mass, following the likelihood analysis of Ref.~\cite{pandax2017}. A standard profile likelihood test statistic was 
determined as a function of WIMP mass and cross section, and compared with that from a large number of toy MC calculations to derive the upper limits of the signal yields at $90\%$ confidence level (C.L.)~\cite{cls1,cls2}.    

\begin{figure}[htbp]
  \centering
  \includegraphics[width=0.45\textwidth]{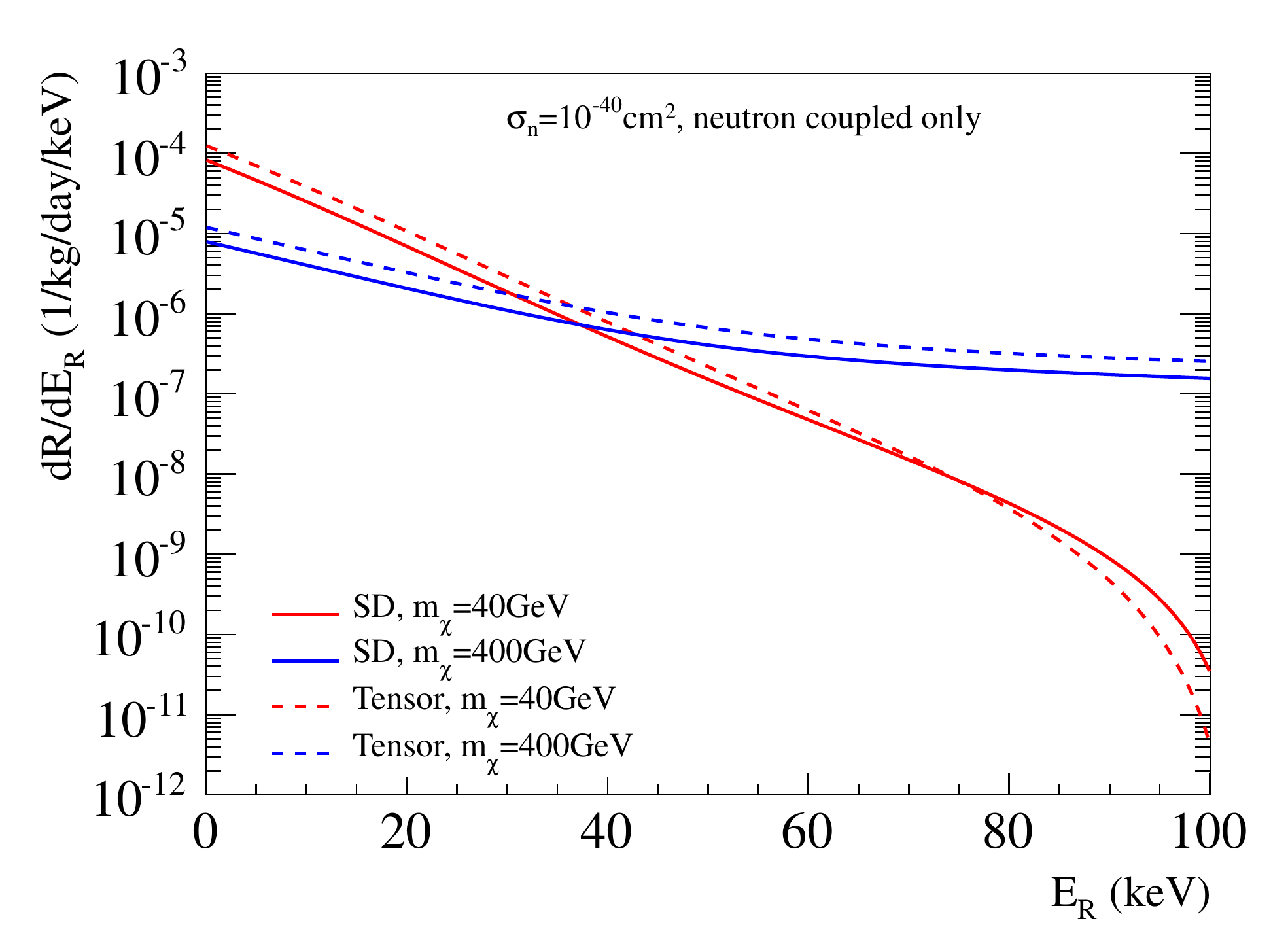}
  \includegraphics[width=0.45\textwidth]{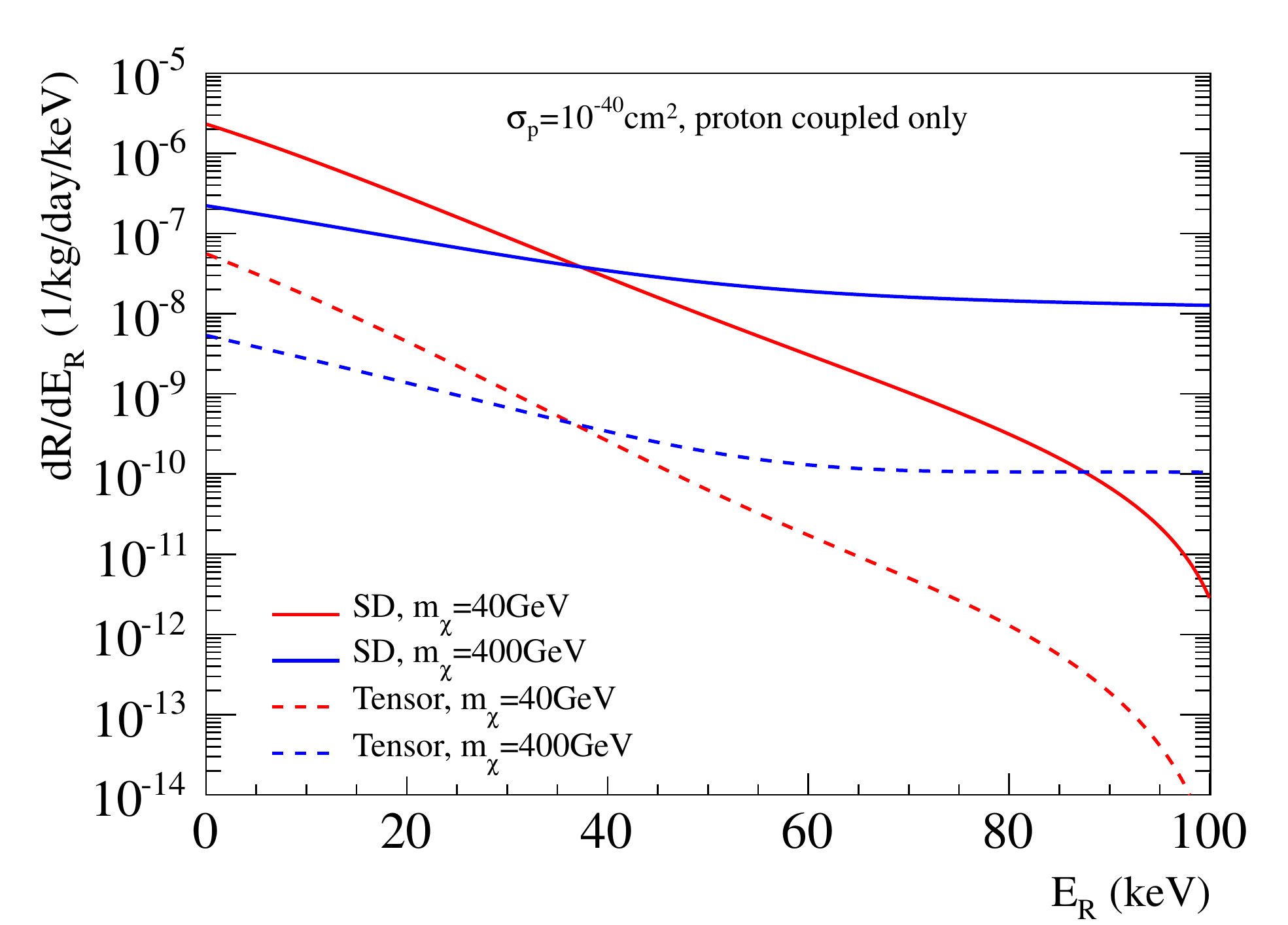}
  \caption{Recoil energy spectra for the scattering of spin-${1 \over 2}$ WIMPs on xenon nuclei, for WIMP masses of ${\rm 40~GeV}/c^2$ and ${\rm 400~GeV}/c^2$,  for SD and tensor interactions.  The top (bottom) panel corresponds to neutron-only (proton-only) coupling.}
  \label{fig:recoilSD}
\end{figure}

The upper panel of Fig.~\ref{fig:EFT} gives PandaX-II bounds on the coherent SI interaction $\mathcal{L}_\mathrm{int}^5$ as well as the spin-spin interaction $\mathcal{L}^{15}_\mathrm{int}$, 
related to the traditional SD interaction discussed below.  They are well below the nominal weak scale $d \sim 1$ over all or almost all WIMP mass range illustrated.  The isoscalar coupling 
limit $|d_5^0| < 5.6 \cdot 10^{-5}$ at maximum sensitivity ($\sim$ 40 GeV WIMP) corresponds to the PandaX-II cross section bound $\sigma^\mathrm{SI}_{p,n} < 8.6 \cdot 10^{-47}$ cm$^2$.  
But limits on the axial (WIMP) - vector (nucleus) coupling $d_{13}^{0,1}$ are also at the weak scale, illustrating the power of current-generation experiments to 
probe momentum- or velocity-dependent interactions.  The lower panel gives the corresponding bounds for interaction involving WIMP magnetic and electric dipole moments.
The bounds are near and in one case (the WIMP electric dipole moment coupling to the nucleon vector current $\mathcal{L}^{17}_\mathrm{int}$) below the nominal weak scale.
Bounds given in the lower panel also include an implicit dimensionless factor representing any needed rescaling of the WIMP moments to their physical values,
relative to the nucleon scale we adopted via $m_M \sim m_N$.

\begin{figure}[htbp]
  \centering
  \includegraphics[width=0.45\textwidth]{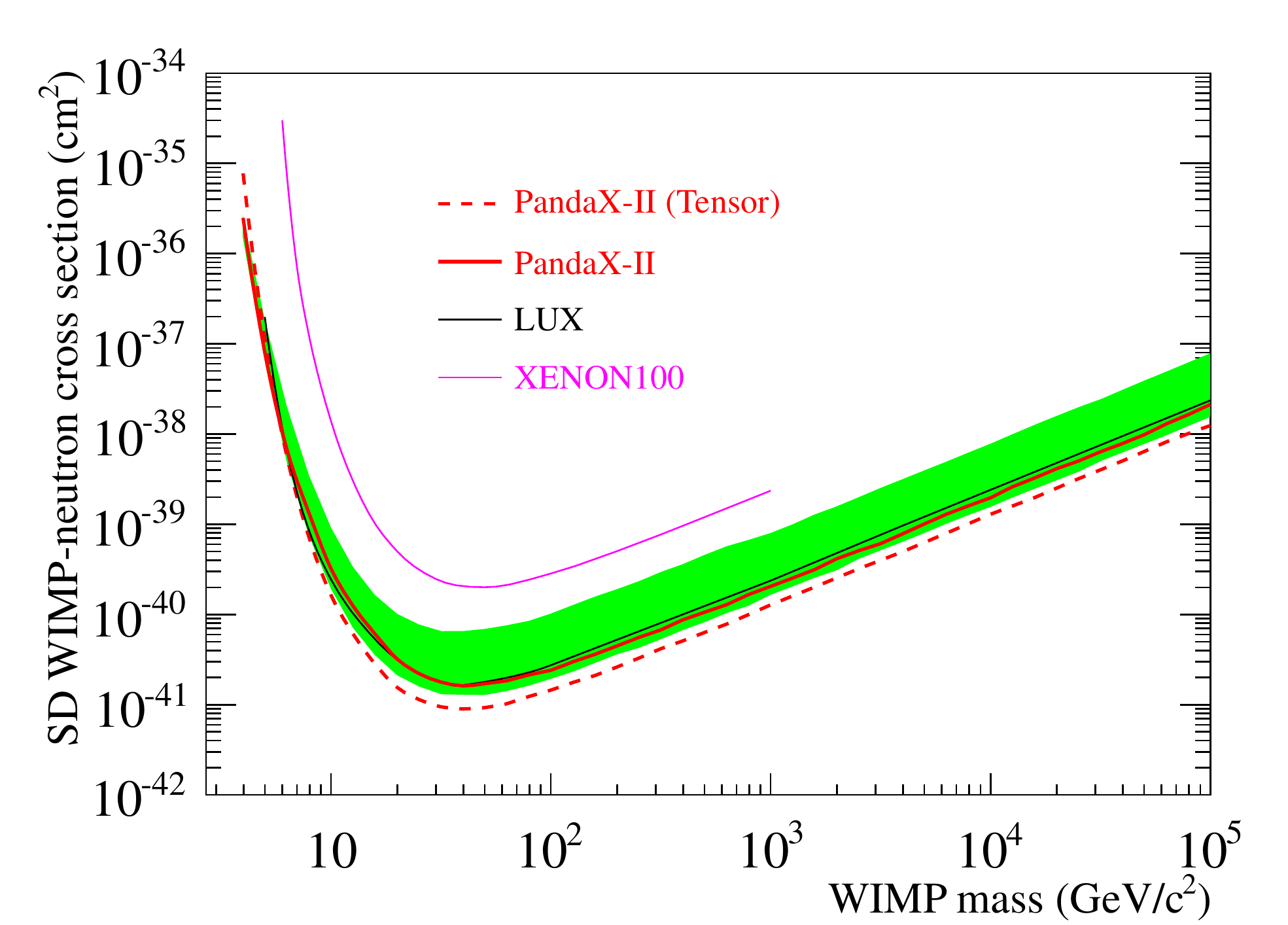}
  \includegraphics[width=0.45\textwidth]{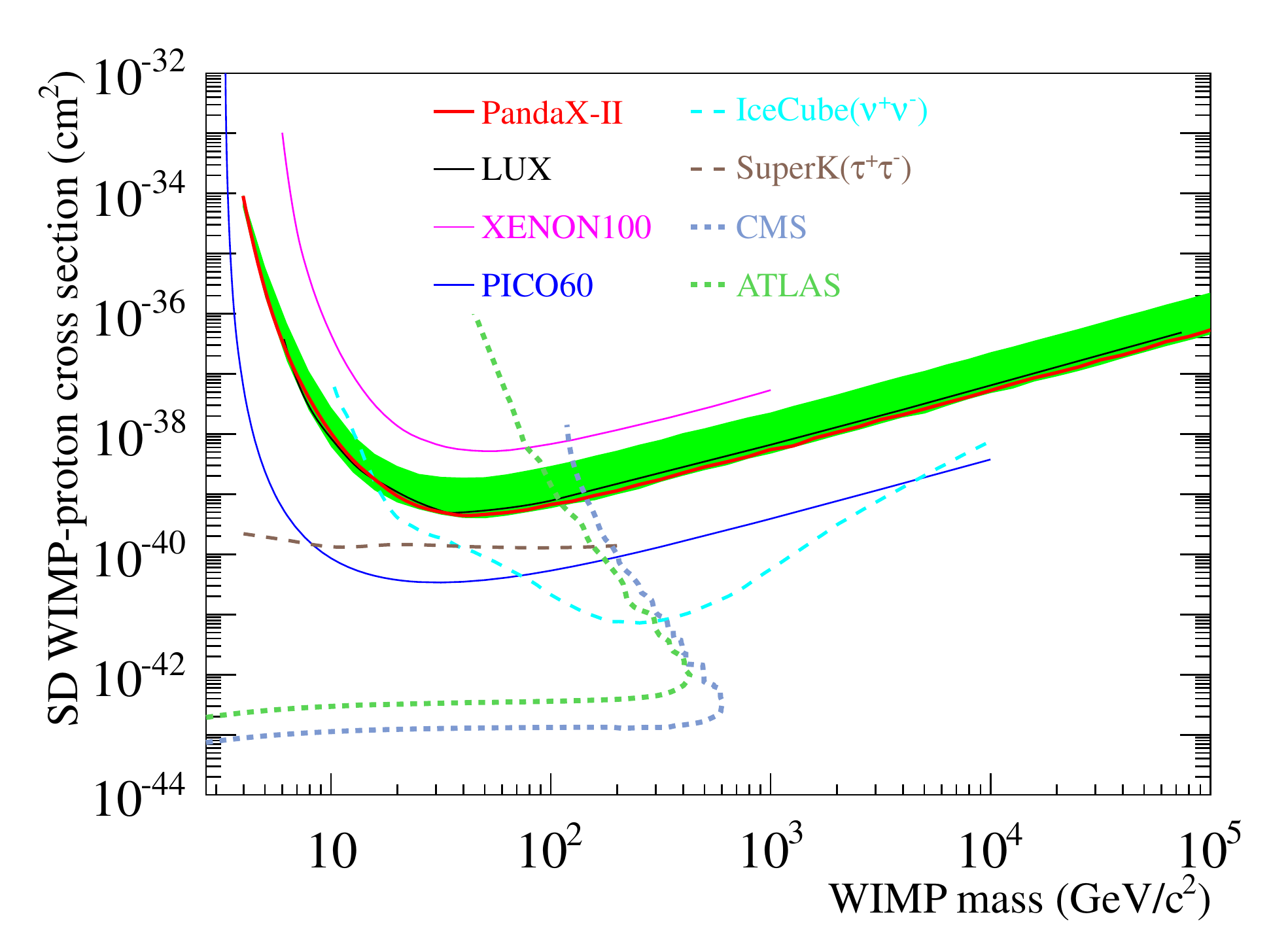}
  \caption{PandaX-II exclusion limits on the WIMP-nucleon cross section for the standard SD interaction assuming neutron-only (top panel) and proton-only (bottom panel)  coupling. 
  The 1$\sigma$ sensitivity bands are shown in green. 
  Also shown are recent results from  LUX~\cite{luxsd2017}, XENON100~\cite{xenonsd}, ATLAS~\cite{atlasmonojet2017}, CMS~\cite{cmsmonojet2017}, PICO-2L~\cite{pico2l_2016},PICO-60~\cite{pico60_2016,pico60_c3f8_2017}, IceCube~\cite{icecube} and Super-K~\cite{superk}. The dashed line (top panel)
 gives PandaX-II limits on tensor WIMP-neutron couplings.}
  \label{fig:SD}
\end{figure}

An often-used measure of experimental sensitivity to spin-dependent WIMP scattering is provided by the SD axial nucleon-level operator obtained from Eq. (1),
\begin{equation}
\mathcal{O}_\mathrm{SD}  \equiv {  g_A(q^2) \over g_A(0)} \mathcal{O}_4 -  {m_N g_P(q^2) \over 2 g_A(0)} \mathcal{O}_6
\end{equation}
with the specific combination of nucleon-level axial and induced pseudoscalar operators dependent on isospin through the $C_q^{AA}$.   The pseudoscalar coupling is enhanced in the
isovector channel by pion-pole dominance.  Comparisons are typically made by selecting proton-only ($p$) or neutron-only ($n$) couplings.  The nuclear cross section $\sigma^\mathrm{SD \, A}_{p,n}$ is related to the nucleon cross 
section $\sigma^\mathrm{SD}_{p,n}$ by
\begin{equation}
\sigma^\mathrm{SD \, A}_{p,n}(q^2) ={4 \pi \over 3(2J+1)} \left({\mu_A \over \mu_{p,n} }\right) ^2 S^A_{p,n}(q^2) \sigma^\mathrm{SD}_{p,n} 
\end{equation}
where $\mu_A$ ($\mu_{p,n}$) is the reduced mass for scattering off the nucleus (nucleon), and $S^A_{p,n}(q^2)$ is the nuclear spin structure function that we take
from \cite{klos2013}, including exchange current corrections.
Figure~\ref{fig:recoilSD} shows recoil spectra for the SD and tensor cross sections.

The $90\%$ C.L. cross section limits are shown in Fig.~\ref{fig:SD}. The optimal bounds, obtained at $M_\mathrm{WIMP} \sim {\rm 40~GeV}/c^2$, limit
$\sigma^\mathrm{SD}_n$ (neutron-only) and $\sigma^\mathrm{SD}_p$ (proton-only) to  $\rm 1.6 \times 10^{-41}$ cm$^2$  and $\rm 4.4\times 10^{-40}$ cm$^2$, respectively.
The results modestly improve existing LUX bounds \cite{lux2017} for $M_\mathrm{WIMP} > {\rm 40~GeV}/c^2$.

Models of asymmetric dark matter favor Dirac fermion WIMPs, where candidate dimension-4 effective operators include the SD and tensor
$\bar{\chi} \sigma_{\mu \nu} \chi ~ \bar{q} \sigma^{\mu \nu} q$ operators.  Constraints from direct detection are particularly competitive for the latter \cite{buckley}.  The tensor interaction generates,
after a leading-order chiral reduction \cite{grinstein2017}, the nucleon-level operator $8 \mathcal{O}_4$.  The dashed curve in the top panel of Fig. \ref{fig:SD} shows the neutron-only PandaX-II limits.  Bounds on $\sigma^\mathrm{T} _n$ and
$\sigma^\mathrm{T} _p$ of
$\rm 9.0\times 10^{-42}\, cm^2$ and $\rm 2.2 \times 10^{-38} \, cm^2$, respectively, are found at $M_\mathrm{WIMP} \sim$ 40 GeV.

 In conclusion, we have presented new limits on a candidate spin-dependent WIMP interactions, using PandaX-II Run~9 and Run~10 data with an exposure of 54-ton-days.  In addition
 to the standard SD interaction, we considered vector-axial vector interferences, interactions generated by WIMP magnetic and electric dipole moments, and tensor interactions.   
 We showed that PandaX-II has achieved sufficient sensitivity to
probe certain velocity- and momentum-dependent interactions at the weak scale.
We obtained the most stringent upper limits to date on $\sigma^\mathrm{SD}_n$ for $M_\mathrm{WIMP}$ above $ {\rm 40~GeV/c}^2$, with a lowest excluded value of $1.6 \times10^{-41}\rm cm^2$ at 40 GeV/c$^2$, 90\% c.l.  The corresponding proton and tensor interaction constraints are $\sigma^\mathrm{SD}_p < 4.4 \times 10^{-40} {\rm cm^2}$, 
$\sigma^\mathrm{T}_n < 9.0 \times 10^{-42} {\rm cm^2}$, and $\sigma^\mathrm{T}_p < 2.2 \times 10^{-38} {\rm cm^2}$.

\section*{Acknowledgments}
This project has been supported by a 985-III grant from Shanghai
Jiao Tong University, grants from National Science Foundation of
China (Nos. 11435008, 11455001, 11505112, 11525522, 11775141 and 11755001), a
grant from the Ministry of Science and Technology of China
(No. 2016YFA0400301).  We thank the Office of Science and Technology, Shanghai Municipal
Government (No. 11DZ2260700, No. 16DZ2260200, No. 18JC1410200) and the Key
Laboratory for Particle Physics, Astrophysics and Cosmology,
Ministry of Education, for important support. This work is supported in part by the Chinese
Academy of Sciences Center for Excellence in Particle Physics
(CCEPP) and Hongwen Foundation in Hong Kong. WH is supported by the US Department of Energy (DE-SC00046548 and
DE-AC02-05CH11231) and the National Science Foundation (PHY-1630782).
We would like to thank Dr. Yue-Lin Sming Tsai and Dr. Zuowei Liu for the useful
discussions on the EFT models in direct detection experiment.
Finally, we thank the CJPL administration and the Yalong River Hydropower Development
Company Ltd. for indispensable logistical support and other help.

\section*{References}
\bibliography{sdEFT}
\end{document}